# A Note on the Objectivity (Rotational Invariance) of the Stored Energy Density in Continuum Physics

Jiashi Yang (jyang1@unl.edu)
Department of Mechanical and Materials Engineering
University of Nebraska-Lincoln, Lincoln, NE 68588-0526, USA

This short note is concerned with the rotational invariance of the stored energy density in continuum physics as a scalar function of a few vectors. A simple derivation is presented for the determination of the general form of the energy density in the case of a two-dimensional space. It is also shown that the general form of the energy density so determined may be further reduced. The three-dimensional case is also discussed.

Objectivity is a fundamentally important concept in continuum physics. It refers to the rotational invariance of physical quantities under time-dependent rotations described by an orthogonal matrix $\mathbf{Q}$. Specifically, we consider the objectivity of the stored energy density which typically is a scalar function of a few vectors such as the deformation gradient $\mathbf{F}$ and the electric as well as magnetic fields. $\mathbf{F}$ is a two-point tensor which is equivalent to three vectors with respect to the spatial coordinate only. There exist several arguments that for rotational invariance the energy density can only depend on $\mathbf{F}$ through the deformation tensor $\mathbf{C}=\mathbf{F}^T \cdot \mathbf{F}$. However, the one extensively used in the literature has been shown to have a logical fallacy [1]. It is due to setting $\mathbf{Q}=\mathbf{R}$, the rotation tensor in the polar decomposition of $\mathbf{F}$, and that $\mathbf{R}$ is a two-point tensor but $\mathbf{Q}$ is not. A few authors [2-4] cited a theorem by Cauchy [5] for objectivity but [5] is difficult to procure. An analytical proof of Cauchy's theory is given in [4] but is has not been widely received. The proof in [4] is for three-dimensional vectors which is somewhat involved. We examine the two-dimensional case below which is rather simple and revealing.

A two-dimensional rotation is described by

$$\mathbf{Q} = \begin{bmatrix} \cos\theta & -\sin\theta \\ \sin\theta & \cos\theta \end{bmatrix}, \quad \mathbf{Q}\cdot\mathbf{Q}^T = \mathbf{1}, \quad \det(\mathbf{Q})=1. \tag{1}$$

Under $\mathbf{Q}$, a vector $\mathbf{v}$ becomes

$$\mathbf{v}' = \mathbf{Q}\cdot\mathbf{v} = \begin{bmatrix} \cos\theta & -\sin\theta \\ \sin\theta & \cos\theta \end{bmatrix}\begin{bmatrix} v_1 \\ v_2 \end{bmatrix} = \begin{bmatrix} v_1\cos\theta - v_2\sin\theta \\ v_1\sin\theta + v_2\cos\theta \end{bmatrix} = \begin{bmatrix} v_1' \\ v_2' \end{bmatrix}. \tag{2}$$

Consider the simplest case of a scalar function $f$ of one two-dimensional vector $\mathbf{v}$ only first. For rotational invariance, $f$ must satisfy

$$f(\mathbf{v}) = f(\mathbf{v}'), \tag{3}$$

or

$$f(v_1,v_2) = f(v_1',v_2'), \tag{4}$$

$$f(v_1,v_2) = f(v_1\cos\theta - v_2\sin\theta, v_1\sin\theta + v_2\cos\theta). \tag{5}$$

Differentiating both sides of Eq. (5) with respect to $\theta$, we obtain

$$0 = \frac{\partial f}{\partial v_1'}(-v_1\sin\theta - v_2\cos\theta) + \frac{\partial f}{\partial v_2'}(v_1\cos\theta - v_2\sin\theta), \tag{6}$$

or

$$-v_2'\frac{\partial f}{\partial v_1'} + v_1'\frac{\partial f}{\partial v_2'} = 0, \tag{7}$$

which is a first-order linear and homogeneous partial differential equation for $f$. Its characteristic

equation is
$$\frac{dv'_1}{-v'_2} = \frac{dv'_2}{v'_1}, \tag{8}$$

or
$$v'_1 dv'_1 + v'_2 dv'_2 = 0. \tag{9}$$

A first integral of Eq. (9) is
$$(v'_1)^2 + (v'_2)^2 = \mathbf{v}' \cdot \mathbf{v}' = C. \tag{10}$$

Then the general solution of Eq. (7) can be written as
$$f = f[(v'_1)^2 + (v'_2)^2] = f(\mathbf{v}' \cdot \mathbf{v}') = f(\mathbf{v} \cdot \mathbf{v}). \tag{11}$$

Similarly, when $f$ is a function of two vectors, $\mathbf{u}$ and $\mathbf{v}$, for rotational invariance,
$$f(\mathbf{u}, \mathbf{v}) = f(\mathbf{u}', \mathbf{v}'), \tag{12}$$

or
$$f(u_1, u_2, v_1, v_2) = f(u'_1, u'_2, v'_1, v'_2), \tag{13}$$

$$f(u_1, u_2, v_1, v_2)$$
$$= f(u_1 \cos\theta - u_2 \sin\theta, u_1 \sin\theta + u_2 \cos\theta, v_1 \cos\theta - v_2 \sin\theta, v_1 \sin\theta + v_2 \cos\theta). \tag{14}$$

Differentiating both sides of Eq. (14) with respect to $\theta$, we obtain
$$0 = \frac{\partial f}{\partial u'_1}(-u_1 \sin\theta - u_2 \cos\theta) + \frac{\partial f}{\partial u'_2}(u_1 \cos\theta - u_2 \sin\theta)$$
$$+ \frac{\partial f}{\partial v'_1}(-v_1 \sin\theta - v_2 \cos\theta) + \frac{\partial f}{\partial v'_2}(v_1 \cos\theta - v_2 \sin\theta), \tag{15}$$

or
$$-u'_2 \frac{\partial f}{\partial u'_1} + u'_1 \frac{\partial f}{\partial u'_2} - v'_2 \frac{\partial f}{\partial v'_1} + v'_1 \frac{\partial f}{\partial v'_2} = 0. \tag{16}$$

The characteristic equations of Eq. (16) are
$$\frac{du'_1}{-u'_2} = \frac{du'_2}{u'_1} = \frac{dv'_1}{-v'_2} = \frac{dv'_2}{v'_1}. \tag{17}$$

The first integrals are
$$(u'_1)^2 + (u'_2)^2 = \mathbf{u}' \cdot \mathbf{u}' = C_1,$$
$$(v'_1)^2 + (v'_2)^2 = \mathbf{v}' \cdot \mathbf{v}' = C_2, \tag{18}$$
$$u'_1 v'_1 + u'_2 v'_2 = \mathbf{u}' \cdot \mathbf{v}' = C_3.$$

Then the general solution of Eq. (16) is
$$f = f(\mathbf{u}' \cdot \mathbf{u}'; \mathbf{v}' \cdot \mathbf{v}'; \mathbf{u}' \cdot \mathbf{v}') = f(\mathbf{u} \cdot \mathbf{u}; \mathbf{v} \cdot \mathbf{v}; \mathbf{u} \cdot \mathbf{v}). \tag{19}$$

Thus $f$ can only be a function of the three inner products of $\mathbf{u}$ and $\mathbf{v}$. When $f$ is a function of three vectors in a two-dimensional space, $\mathbf{u}$, $\mathbf{v}$ and $\mathbf{w}$, we have
$$-u'_2 \frac{\partial f}{\partial u'_1} + u'_1 \frac{\partial f}{\partial u'_2} - v'_2 \frac{\partial f}{\partial v'_1} + v'_1 \frac{\partial f}{\partial v'_2} - w'_2 \frac{\partial f}{\partial w'_1} + w'_1 \frac{\partial f}{\partial w'_2} = 0. \tag{20}$$

The characteristic equations of Eq. (20) are
$$\frac{du'_1}{-u'_2} = \frac{du'_2}{u'_1} = \frac{dv'_1}{-v'_2} = \frac{dv'_2}{v'_1} = \frac{dw'_1}{-w'_2} = \frac{dw'_2}{w'_1}. \tag{21}$$

The following six first integrals can be found:

$$(u_1')^2 + (u_2')^2 = \mathbf{u}' \cdot \mathbf{u}' = C_1,$$
$$(v_1')^2 + (v_2')^2 = \mathbf{v}' \cdot \mathbf{v}' = C_2, \qquad (22)$$
$$(w_1')^2 + (w_2')^2 = \mathbf{w}' \cdot \mathbf{w}' = C_3,$$

and

$$u_1'v_1' + u_2'v_2' = \mathbf{u}' \cdot \mathbf{v}' = C_4,$$
$$u_1'w_1' + u_2'w_2' = \mathbf{u}' \cdot \mathbf{w}' = C_5, \qquad (23)$$
$$v_1'w_1' + v_2'w_2' = \mathbf{v}' \cdot \mathbf{w}' = C_6.$$

Then the general solution of Eq. (20) can be written as

$$f = f(\mathbf{u}' \cdot \mathbf{u}'; \mathbf{v}' \cdot \mathbf{v}'; \mathbf{w}' \cdot \mathbf{w}'; \mathbf{u}' \cdot \mathbf{v}'; \mathbf{u}' \cdot \mathbf{w}'; \mathbf{v}' \cdot \mathbf{w}')$$
$$= f(\mathbf{u} \cdot \mathbf{u}; \mathbf{v} \cdot \mathbf{v}; \mathbf{w} \cdot \mathbf{w}; \mathbf{u} \cdot \mathbf{v}; \mathbf{u} \cdot \mathbf{w}; \mathbf{v} \cdot \mathbf{w}). \qquad (24)$$

We note that three vectors in a two-dimensional space are not linearly independent. As a consequence, only five of the six first-integrals in Eqs. (22) and (23) are independent. This can be seen as follows. Let

$$\mathbf{w} = \alpha \mathbf{u} + \beta \mathbf{v}. \qquad (25)$$

Dotting both sides of Eq. (25) by $\mathbf{u}$ and $\mathbf{v}$, respectively, we have

$$(\mathbf{w} \cdot \mathbf{u}) = \alpha(\mathbf{u} \cdot \mathbf{u}) + \beta(\mathbf{v} \cdot \mathbf{u}),$$
$$(\mathbf{w} \cdot \mathbf{v}) = \alpha(\mathbf{u} \cdot \mathbf{v}) + \beta(\mathbf{v} \cdot \mathbf{v}). \qquad (26)$$

Equation (26) determines $\alpha$ and $\beta$ in terms of the five inner products in Eq. (26). Then

$$\mathbf{w} \cdot \mathbf{w} = (\alpha \mathbf{u} + \beta \mathbf{v}) \cdot (\alpha \mathbf{u} + \beta \mathbf{v}), \qquad (27)$$

which can be expressed by the five inner products in Eq. (26).

Finally, for convenience and completeness, we present the result for a scalar function of three-dimensional vectors [4] below. For a scalar function $f$ of $N$ three-dimensional vectors to be rotationally invariant, $f$ must satisfy

$$f(v_i^{(1)}, \cdots, v_i^{(n)}, \cdots, v_i^{(N)}) = f(v_i'^{(1)}, \cdots, v_i'^{(n)}, \cdots, v_i'^{(N)})$$
$$= f(Q_{ij} v_j^{(1)}, \cdots, Q_{ij} v_j^{(n)}, \cdots, Q_{ij} v_j^{(N)}). \qquad (28)$$

The differentiation of both sides of Eq. (28) with respect to $Q_{pq}$ leads to

$$0 = \frac{\partial f}{\partial Q_{pq}} dQ_{pq}. \qquad (29)$$

Since the components of $\mathbf{Q}$ are not independent, i.e., $\mathbf{Q}$ is orthogonal with the following constraint:

$$Q_{mk} Q_{nk} = \delta_{mn}, \qquad (30)$$

we construct

$$F(\mathbf{Q}) = f(Q_{ij} v_j^{(1)}, \cdots, Q_{ij} v_j^{(n)}, \cdots, Q_{ij} v_j^{(N)}) - \frac{1}{2} \lambda_{mn} (Q_{mk} Q_{nk} - \delta_{nm}),$$
$$\lambda_{mn} = \lambda_{nm}, \qquad (31)$$

where $\lambda_{mn}$ are Lagrange multipliers. The differentiation of $F$ with respect to $Q_{pq}$ leads to

$$0 = \sum_{n=1}^{N} \frac{\partial f}{\partial v_p'^{(n)}} v_q^{(n)} - \lambda_{mp} Q_{mq}. \qquad (32)$$

Multiplying Eq. (32) by $Q_{rq}$, we obtain

$$0 = \sum_{n=1}^{N} \frac{\partial f}{\partial v'^{(n)}_p} v^{(n)}_q Q_{rq} - \lambda_{mp} Q_{mq} Q_{rq}, \tag{33}$$

or

$$\sum_{n=1}^{N} \frac{\partial f}{\partial v'^{(n)}_p} v'^{(n)}_r = \lambda_{rp}. \tag{34}$$

Since $\lambda_{rp}$ is symmetric, Eq. (34) implies that

$$\sum_{n=1}^{N} \frac{\partial f}{\partial v'^{(n)}_p} v'^{(n)}_r = \sum_{n=1}^{N} \frac{\partial f}{\partial v'^{(n)}_r} v'^{(n)}_p. \tag{35}$$

Equation (35) represents nine first-order partial differential equations for *f*. Only three of them are nontrivial and independent. Let the inner products among the vectors be

$$C^{(rs)} = v^{(r)}_k v^{(s)}_k, \quad r,s = 1,2,\cdots,N. \tag{36}$$

It can be verified that

$$f = f(C^{(11)}, C^{(12)}, \cdots, C^{(rs)}, \cdots C^{(NN)}) \tag{37}$$

satisfies Eq. (35).